\documentclass[pre,twocolumn,showpacs]{revtex4}
\pdfoutput=1
\usepackage{amssymb,latexsym}
\usepackage{amsmath,amsbsy,bbm}
\usepackage{ifpdf}
\usepackage{epsfig,bm}
\usepackage{graphicx,comment}
\usepackage{color}
\usepackage{soul}
\usepackage[normalem]{ulem}
\begin{document}

\title{A Neural Networks study of the phase transitions of Potts model}
\author{D.-R. Tan}
\affiliation{Department of Physics, National Taiwan Normal University,
88, Sec.4, Ting-Chou Rd., Taipei 116, Taiwan}
\author{C.-D. Li}
\affiliation{Department of Physics, National Taiwan Normal University,
88, Sec.4, Ting-Chou Rd., Taipei 116, Taiwan}
\author{W.-P. Zhu}
\affiliation{Department of Physics, National Taiwan Normal University,
88, Sec.4, Ting-Chou Rd., Taipei 116, Taiwan}
\author{F.-J. Jiang}
\email[]{fjjiang@ntnu.edu.tw}
\affiliation{Department of Physics, National Taiwan Normal University,
88, Sec.4, Ting-Chou Rd., Taipei 116, Taiwan}

\begin{abstract}
Using the techniques of Neural Networks (NN), we study 
the three-dimensional (3D) 5-state ferromagnetic Potts model
on the cubic lattice as well as the two-dimensional (2D) 
3-state antiferromagnetic Potts model on the square lattice.
Unlike the conventional approach, here
we follow the idea employed in Ann.~Phy.~391 (2018) 312-331. 
Specifically, instead of numerically generating numerous objects 
for the training, the whole or 
part of the theoretical ground state configurations
of the studied models are considered as the training sets. 
Remarkably,
our investigation of these two models provides convincing
evidence for the effectiveness of the method of preparing
training sets used in this study. In particular,
the results of the 3D model obtained here imply that the NN
approach is as efficient as the traditional method since
the signal of a first order phase transition, namely
tunneling between two channels, determined by the
NN method is as strong as that calculated with the Monte Carlo
technique. Furthermore,
the outcomes associated with the considered 2D system
indicate even little partial information of the ground states
can lead to conclusive results regarding the studied phase 
transition. The achievements reached in our investigation
demonstrate that the performance of NN, using certain amount of the theoretical
ground state configurations as the training sets, is impressive.  
  
\end{abstract}


\maketitle

\section{Introduction}

During the last couple years, the application of methods and techniques
of artificial intelligence (AI) in many-body systems has drawn tremendous
attention in the physics community \cite{Rup12,Sny12,Mer14,Sch14,Li15,Lee16,Wan16,Oht16,Car16,Tro16,Tor16,Bro16,Chn16,Tan16,Nie16,Liu16,Xu16,Wan17,Liu17,Che17,Nag17,Den17,Pon17,Zha17,Zha17.1,Hu17,Li18,Chn18,Bea18,Sha18,Zha18,Gao18,Zha19,Gre19,Don19,Yoo19,Can19,Lia19}. For example, by employing
the idea of restricted Boltzmann machine, it is demonstrated
that the efficiency of certain Monte Carlo simulations can be
improved dramatically \cite{Wan17}. In addition, with the supervised and unsupervised
Neural Networks (NN), the critical points and exponents, as well as the nature
of the phase transitions of some classical and quantum models are determined
with high accuracy \cite{Wan16,Oht16,Car16,Tro16,Tor16,Bro16,Chn16,Nie16,Hu17,Li18}. These applications of AI in physics 
are very successful. Hence it is anticipated 
that the ideas of AI not only provide alternative approaches for 
studying many-body systems, but also have great potential in 
exploring properties of materials that are beyond what have been achieved 
using the traditional methods.     

The standard (conventional) procedure of applying supervised NN
to investigate the phase transitions of physics systems 
consists of three steps, namely the training, the validation,
and the testing stages. Taking two-dimensional (2D) Ising model on the square lattice 
as an example \cite{Car16}, in the testing stage, 
typical configurations at various temperatures below and above the transition 
temperature $T_c$ are generated by Monte Carlo simulations
or other numerical techniques. Moreover, labels of $(1,0)$ and $(0,1)$ are 
assigned
to all the generated configurations below and above $T_c$, respectively. 
Through the optimization
procedure, the desired weights are determined and are used in later computations 
in both the validation and testing stages. The role of the 
validation stage is to make sure that correct outcomes are obtained
using the trained NN (weights). Finally, in the testing stage, output
results at many temperatures acrossing $T_c$ are determined. In particular,
the temperature at which the output is $(0.5,0.5)$ is expected to be
the $T_c$. 

Using the procedures described in the previous paragraph, it is demonstrated
that the $T_c$ of 2D Ising model on the square lattice indeed can be calculated 
accurately using a supervised NN \cite{Car16}. Furthermore, NN can even detect incorrect
information and precisely determine $T_c$ \cite{Nie16}. Such a conventional approach also applies 
to other models and success to certain satisfactory extent are obtained. 

While it seems promising that in the near future, methods of AI may play an important role
in studying many-body systems, when it comes to examine the critical phenomena, what are the benefits 
of using the NN techniques rather than 
employing the traditional methods needs further investigation. 
In particular, it is crucial to explore which of the traditional 
and the NN approaches performs better. 
Besides, the conventional 
strategy for the training stage
introduced above has a caveat, namely $T_c$ is known in prior before making a 
use of NN. As a result, for a new system
without the knowledge of its critical point, it may be difficult to employ the
conventional approach to train the NN in a straightforward manner. 

To overcome this issue mentioned above regarding studying a phase transition with a unknown critical point, 
instead of generating configurations numerically
for the training, 
in Ref.~\cite{Li18} the expected ground state configurations are used as the 
training sets of a NN investigation for
the phase transitions of 2D 
$Q$-state ferromagnetic Potts models on the square lattice \cite{Wu82}. Using
this strategy, $T_c$ is not essential in using the NN method
and there is very little computation effort required for generating the training sets. 
With such an unconventional approach,
success of calculating the associated $T_c$ and determining the 
nature of the phase transitions of 2D $Q$-state ferromagnetic Potts models 
are reached \cite{Li18}. 

Although one can locate $T_c$ and determine the nature of phase
transitions accurately for $Q$-state ferromagnetic Potts models with the idea 
of using the theoretical ground state configurations
as the training sets,   
it should be pointed out that for a given positive integer $Q$, there are $Q$ ground state configurations
for $Q$-state ferromagnetic Potts model
and all these configurations can be used 
as the training set like that being done in Ref.~\cite{Li18}, without 
encountering any technical difficulty. An interesting question
arise regarding the applicability of the this approach. Specially,
if only part of all the ground state configurations are employed
as the training set, will the resulting NN still be able to reach
the success as that shown in Ref.~\cite{Li18}? This is an important
effect to examine when systems with highly degenerated ground states
are studied using the NN techniques.  

In addition to studying critical phenomena, the application of AI methods in
the majority fields of science requires the use of real data points as 
the training sets. Indeed, such a combination advances certain areas of
research greatly. Still, it will be extremely compelling to understand 
whether solely AI techniques can achieve the same level of success as that 
obtained by the traditional methods. 

Motivated by these subtle issues described above, here we consider 
NN which are trained without using any actual data as the training sets.
Furthermore, we employ the built NN to study the  
three-dimensional (3D) 5-state ferromagnetic Potts model on the cubic lattice
as well as the 2D 3-state  antiferromagnetic Potts model on the square 
lattice. The reasons that these two models are chosen will be explained
later.

Interestingly, our study for the 3D model indicates that NN is as efficient
as the traditional Monte Carlo method since the signal of a first order phase 
transition, namely tunneling between two channels, determined by the
NN method is as strong as that calculated with the Monte Carlo technique.
This result suggests that NN is a promising alternative approach for studying
many-body systems. Furthermore, the NN outcomes obtained for the considered
2D system provide convincing evidence that using the ideas considered in 
Ref.~\cite{Li18}, even little partial information of the ground states 
can lead to conclusive results regarding the studied phase transition.          
To summarize, the performance of NN, using certain amount of the theoretical
ground state configurations as the training set, is impressive.  

This paper is organized as follows. After the introduction,
the studied microscopic models and the details of the employed NN are 
described. In particular, the NN training sets and labels are introduced 
thoroughly.
Following this the resulting numerical results by applying
the Monte Carlo simulations and the NN techniques are presented. 
Finally, a section concludes our investigation.

\section{The microscopic models and observables}

The Hamiltonian $H$ of $Q$-state Potts model considered in our study is
given by \cite{Wu82,Swe87,Wan89,Wan90}
\begin{equation}
\beta H = -J\beta \sum_{\left< ij\right>} \delta_{\sigma_i,\sigma_j},
\label{eqn1}
\end{equation}
where $\beta$ is the inverse temperature and $\left< ij \right>$ stands for 
the nearest neighboring sites $i$ and $j$. In addition, in Eq.~(\ref{eqn1})
the $\delta$ refers to
the Kronecker function and finally, the Potts variable 
$\sigma_i$ appearing above at each site $i$ takes an integer value from 
$\{1,2,3,...,Q\}$. The situations of $J > 0$ and $J < 0$ correspond to
ferromagnetic and antiferromagnetic Potts models, respectively. 

As already being mentioned previously, in this study we focus on investigating 
the phase transitions of 3D 5-state ferromagnetic Potts model on the cubic 
lattice and 2D 3-state antiferromagnetic Potts model on the square lattice.
The motivations for considering these two models are as follows. 

First of all, it is known that the phase transition of 3D 5-state ferromagnetic Potts model
on the cubic lattice is first order \cite{Wu82}. Furthermore, the signal of a first order
phase transition becomes exponentially hard to observe as the space-time
volume increases \cite{Bil95,Non15}. Therefore, studying 3D 5-state ferromagnetic Potts
model on the cubic lattice provides an opportunity to compare the
efficiency of detecting a first order phase transition 
between the traditional and NN approaches. 

The 2D 3-state antiferromagnetic 
Potts model on the square lattice is studied here because it is shown that its
associated phase transition occurs
at zero temperature \cite{Wan90,Kun95,Fer99}. In other words, the system is disordered at any $T>0$.
As a result, the conventional training strategy usually employed 
in a NN investigation of a many-body system may not
be applicable for this model. Hence the 2D 3-state antiferromagnetic
Potts model on the square lattice serves as a good testing ground
for the NN approach of using the theoretical ground state configurations
as the training set.
   
The obserbables considered here for the 3D 5-state ferromagnetic Potts model
are the energy density $E$ and the magnetization density
$\langle |m| \rangle$. Here $m$ is defined as
\begin{equation}
m = \frac{1}{L^3} \sum_{j} \exp\left(i\frac{2\pi \sigma_j}{5}\right),
\end{equation}
where $L$ is the linear box sizes used in the calculations and the summation
is over all lattice sites $j$.
Moreover, to study
the 2D 3-state antiferromagnetic Potts model on the square lattice,
the staggered magnetization density $m_s$, which takes the form
\begin{eqnarray}
m_s = \frac{1}{3}\sum_{i=1}^{3} |M_i|,
\end{eqnarray}
is measured in our simulations. Here $M_i$ is defined as
\begin{eqnarray}
M_i = \frac{2}{L^2}\sum_x (-1)^{x_1 +x_2}\delta_{\sigma_x,i}, 
\end{eqnarray}
where again the summation is over all lattice sites $x$.
Finally, Potts configurations for both the
considered models are recorded as well and will be used in 
the calculations related to NN.

\section{The constructed supervised Neural Networks}

In this section, we will introduce the details of the supervised NN used in our study. 
The employed training sets and the associated labels will 
be described as well. Moreover, we will consider the simplest NN 
of deep learning and examine whether it can reach the same level of success as those obtained with complicated NN
such as the convolutional Neural Networks (CNN).

\subsection{The built multilayer perceptron (MLP)}

\begin{figure}
\begin{center}
\includegraphics[width=0.5\textwidth]{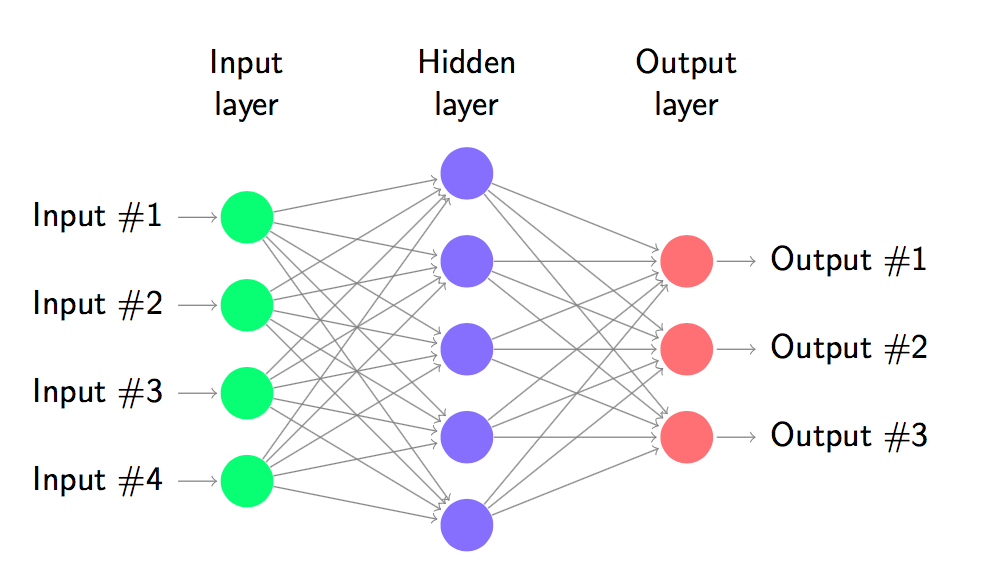}
\end{center}\vskip-0.7cm
\caption{The NN (MLP) used in this study.}
\label{MLP}
\end{figure}

Since we would like to understand whether the simplest deep learning NN (multilayer perceptron, MLP)
is capable of detecting the critical point, the supervised NN considered in our 
investigation consists of only one input layer, one hidden layer of 512 
independent nodes, and one output 
layer using the publicly available NN libraries keras and tensorflow
\cite{kera,tens}. Fig.~\ref{MLP} demonstrates 
the NN used here. The algorithm, optimizer, and
loss function we 
employ for the calculations are minibatch, adam, and categorical cross
entropy, respectively. $L_2$ regularization is applied as well to avoid overfitting.
The activation functions considered are 
ReLU (between the input layer and hidden layer)
and softmax (between the hidden layer and output layer). 
In addition, for the 3D model, computations using various batchsize, 
nodes, copies of the pre-training set (defined later), 
and epoch are conducted as well.
Moreover, the weights obtained in the training processes which minimize the loss function are recorded and are used in later calculations. 
Finally, to understand the impact on the output results from the initial 
values of weights as well as other steps performed in the training stage, 
several sets of random seeds are used
in the investigation. For the studied 2D antiferromagnetic model, 
all the outcomes obtained with various random seeds will be considered in 
determining the final results associated with this model. 

\subsection{Training set and output labels for the 3D model}

\begin{figure}
\begin{center}
\includegraphics[width=0.5\textwidth]{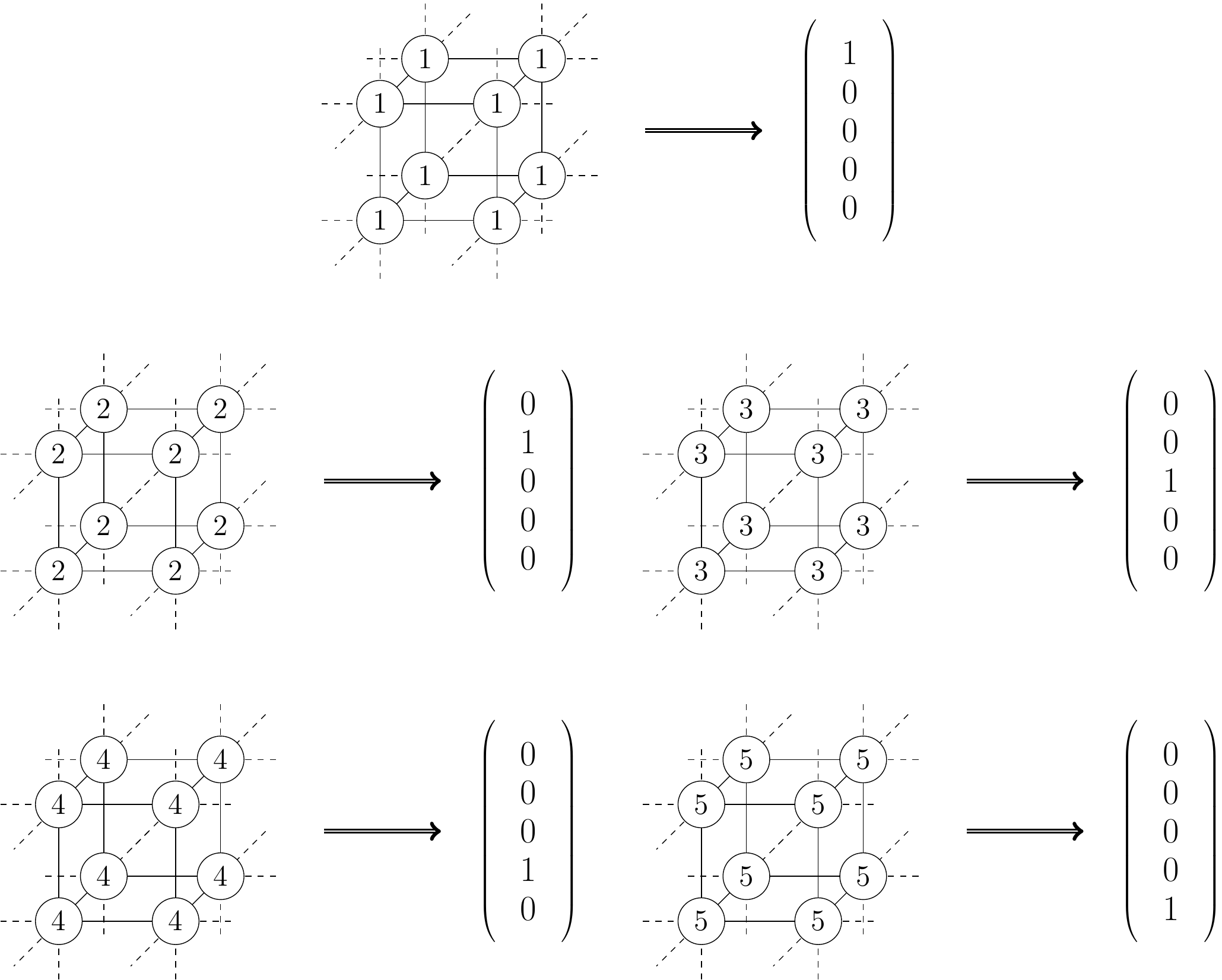}
\end{center}\vskip-0.7cm
\caption{Pre-training set and their corresponding labels considered here for the
3D 5-state ferromagnetic Potts model on the cubic lattice.}
\label{Q53Dlabel}
\end{figure}

Regarding the training set used for the 3D 5-state ferromagnetic Potts model 
on the cubic lattice, we will follow the idea considered in Ref.~\cite{Li18}, namely
the employed training set consists of 200 (or any suitable number) copies of the corresponding
theoretical ground state configurations. The expected ground state 
configurations for 3D 5-state ferromagnetic Potts model on a $L \times L\times L$ 
cubic lattice are obtained by letting the Potts variables on all the lattice sites
take the same (positive) integer from $\{1,2,3,4,5\}$ as their values. Consequently,
there are 5 ground state configurations. The associated labels for these 5
ground state configurations are the basis vectors of five-dimensional (5D)
Euclidean space. While not being unique, clearly one can construct an one to one correspondence 
between the 5 ground state configurations and the basis vectors of 5D Euclidean
space. One of such correspondence is shown in fig.~\ref{Q53Dlabel}. These five
ground state configurations will be named pre-training set in this study.
Finally, we would like to emphasize the fact that when constructing the pre-training set, 
all the allowed Potts variables should be used.

\subsection{The expected output vectors for the 3D model at various $T$}

With such a set up of pre-training set, it is expected
that at extremely low temperatures, the norm ($R$) of the NN output vectors 
are around 1 since most of the Potts variables take the same positive integer $Q_1$
as their values. As a result, one component of the associated
output vector has much larger magnitude than that of the others. 
The norm of such a vector clearly is around 1. 

\begin{figure}
\begin{center}
\includegraphics[width=0.35\textwidth]{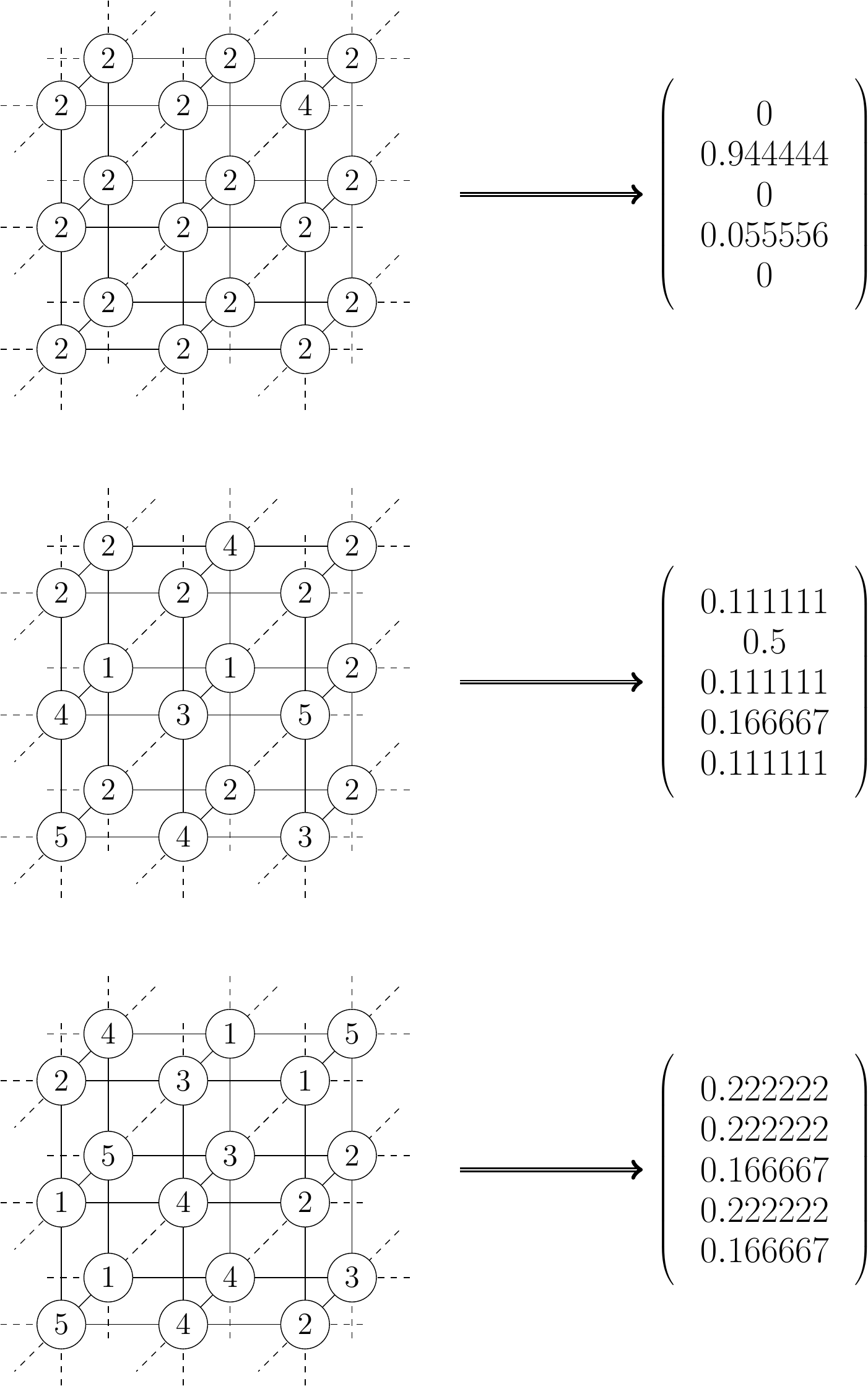}
\end{center}\vskip-0.7cm
\caption{The expected Potts configurations and the corresponding
NN output vectors at low temperatures ($T \ll T_c$, top panel), 
moderate temperature ($T < T_c$, middle panel), and high temperature
($T \ge T_c$, bottom panel) for the studied 3D model. The 
output vectors are for demonstration, not the real ones.}
\label{expected3D}
\end{figure}

As the temperature arises, some Potts variables
begin to obtain other positive integers than $Q_1$. Consequently, the magnitude 
of norm of the output vectors diminishes with $T$. 
When $T\ge T_c$, 
the norm of NN output vectors are around its minimum value $1/\sqrt{5}$. 
This is because
there is an equal probability that each integer in $\{1,2,3,4,5\}$ is the value 
of any Potts variables. The cartoon plots shown in fig.~\ref{expected3D} demonstrates how the Potts 
configurations and the corresponding output vectors change with $T$.
Based on this scenario of $R$ versus $T$, $T_c$ 
can be estimated to lie within the temperature window at which $R$ decreases
rapidly from 1 to $1/\sqrt{5}$. Indeed such a method is shown to be able
to determine $T_c$ with high precision in Ref.~\cite{Li18}. For a more detailed 
introduction to this approach of using the theoretical ground state
configurations as the training set, including its validation, we refer the readers to Ref.~\cite{Li18}.

\subsection{Training set and output labels for the 2D model}

\begin{figure}
\begin{center}
\includegraphics[width=0.5\textwidth]{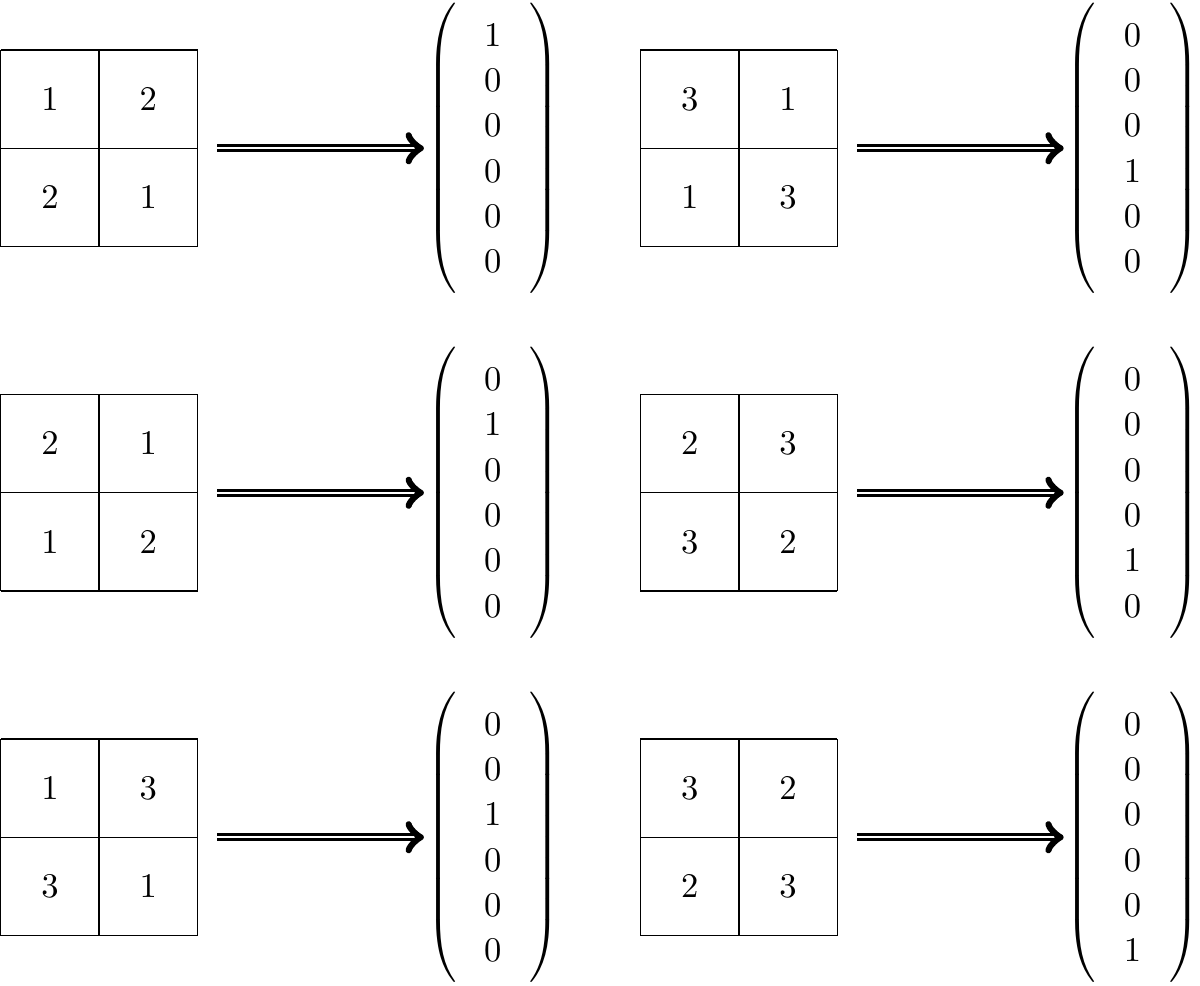}
\end{center}\vskip-0.7cm
\caption{Unit blocks (2 by 2 lattices and their Potts variables) for building the pre-training set consisiting
of 6 configurations, and 
their corresponding labels considered here for the
2D 3-state antiferromagnetic Potts model on the square lattice.}
\label{Q32Dlabel6}
\end{figure}

Similar to the strategy introduced in the previous subsection,
here we will employ the expected ground state 
configurations as the pre-training set for the NN study associated with
the considered 2D system. Unlike the ferromagnetic Potts model, 
any two nearest neighboring Potts variables for the ground state configurations of 2D $Q$-state
antiferromagnetic Potts model differ from each other. As a result, there are tremendous number 
of such configurations when $Q \ge 3$. We construct 6, 18, and 36 expected ground state
configurations of the 2D 3-state antiferromagnetic Potts model and use these configurations
as the pre-training sets. The unit blocks (2 by 2 lattices and their
Potts variables) for the built 6 configurations are shown in
fig.~\ref{Q32Dlabel6}, and configurations on larger lattices are obtained by 
multiplying any of these 6 unit blocks by itself several times in both
the $x$- and $y$-direction (The associated labels which will be introduced shortly 
are demonstrated in fig.~3 as well).
Using these pre-training sets, the actual training sets are
a multiple copy (Here we use a factor of 200) of the pre-training sets. 

\begin{figure}
\begin{center}
\hbox{
~~~~~~~~~~~~~~~~~
\includegraphics[width=0.1\textwidth]{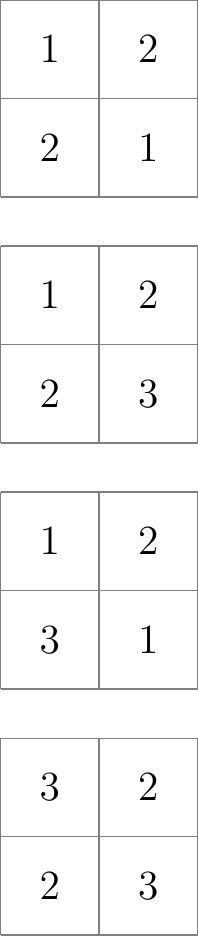}~~~~~~~
\includegraphics[width=0.1\textwidth]{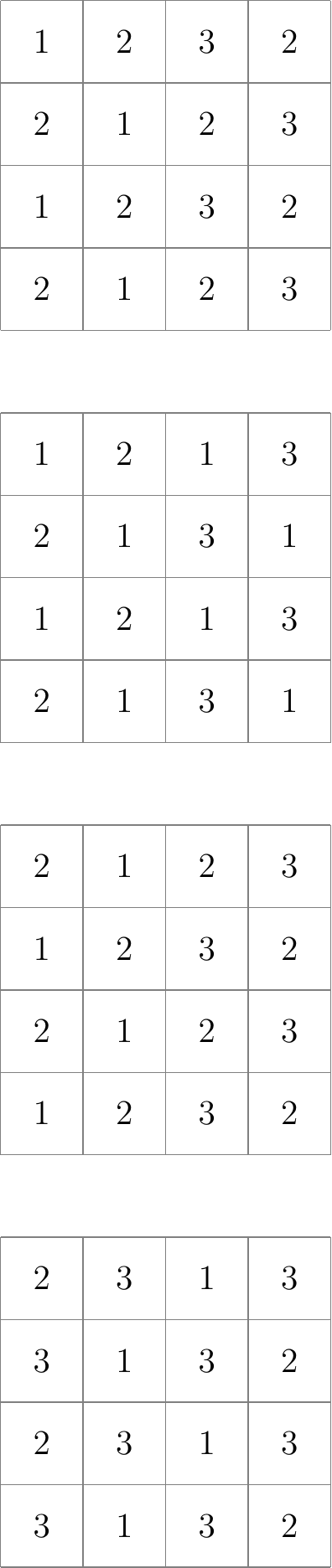}
}
\end{center}\vskip-0.7cm
\caption{Several unit blocks for building pre-training sets consisiting of 18 (left, 2 by 2 lattices and their Potts variables) and 
36 (right, 4 by 4 lattices and their Potts variables) configurations, for the
2D 3-state antiferromagnetic Potts model on the square lattice. Configurations on larger lattices are obtained by 
multiplying any of these unit blocks by itself several times in both the $x$- and $y$-direction.}
\label{Q32Dlabel1836}
\end{figure}

The output labels used here for the pre-training sets follows exactly the ones in the previous  
subsection related to the 3D 5-state ferromagnetic Potts model.
For instance, for the pre-training set consisting of 6 configurations, the corresponding
labels are the basis vectors of 6-dimensional Euclidean space. Clearly, 
similar to the case of ferromagnetic Potts model, one can map 
these 6 configurations in the pre-training set onto the 6 basis vectors of
6-dimensional Euclidean space in an one to one manner (This map is not unique as well).
The same construction rule applies when 18 or 36 configurations are
considered as the pre-training set, see fig.~\ref{Q32Dlabel1836} 
for part of the pre-training sets consisting of 18 and 36 configurations.

Remarkably, although in this study only very little information of the whole ground state configurations
are employed as the training sets, as we will demonstrate shortly, 
the built NN with the designed training sets and labels is
capable of showing convincing evidence that the phase transition of the investigated
2D 3-state antiferromagnetic Potts model occurs only at zero temperature as the theory predicts.

\section{Numerical results}

To generate configurations of the studied 2D and 3D Potts models,
which will be used in the testing stages of the NN procedures, Swendsen and Wang algorithm \cite{Swe87}, Wolff algorithm \cite{Wol89}, 
as well as Swendsen-Wang-Kotecky algorithm \cite{Wan89,Wan90} are adopted.
Particularly, Potts configurations are stored once in every thousand (or two 
thousand) Monte Carlo sweeps after the thermalization. 

\begin{figure}
\begin{center}
\vbox{
\includegraphics[width=0.35\textwidth]{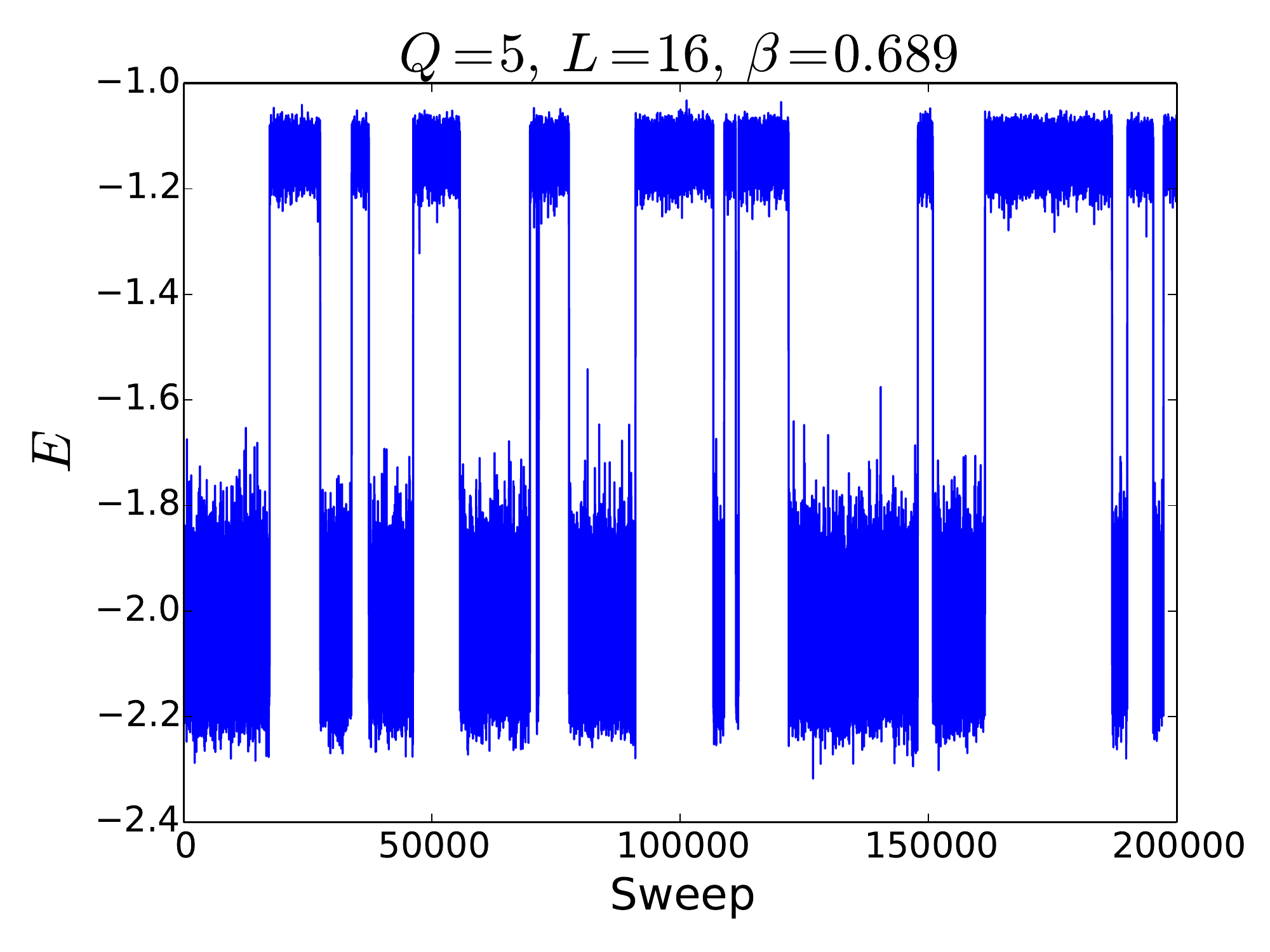}
\includegraphics[width=0.35\textwidth]{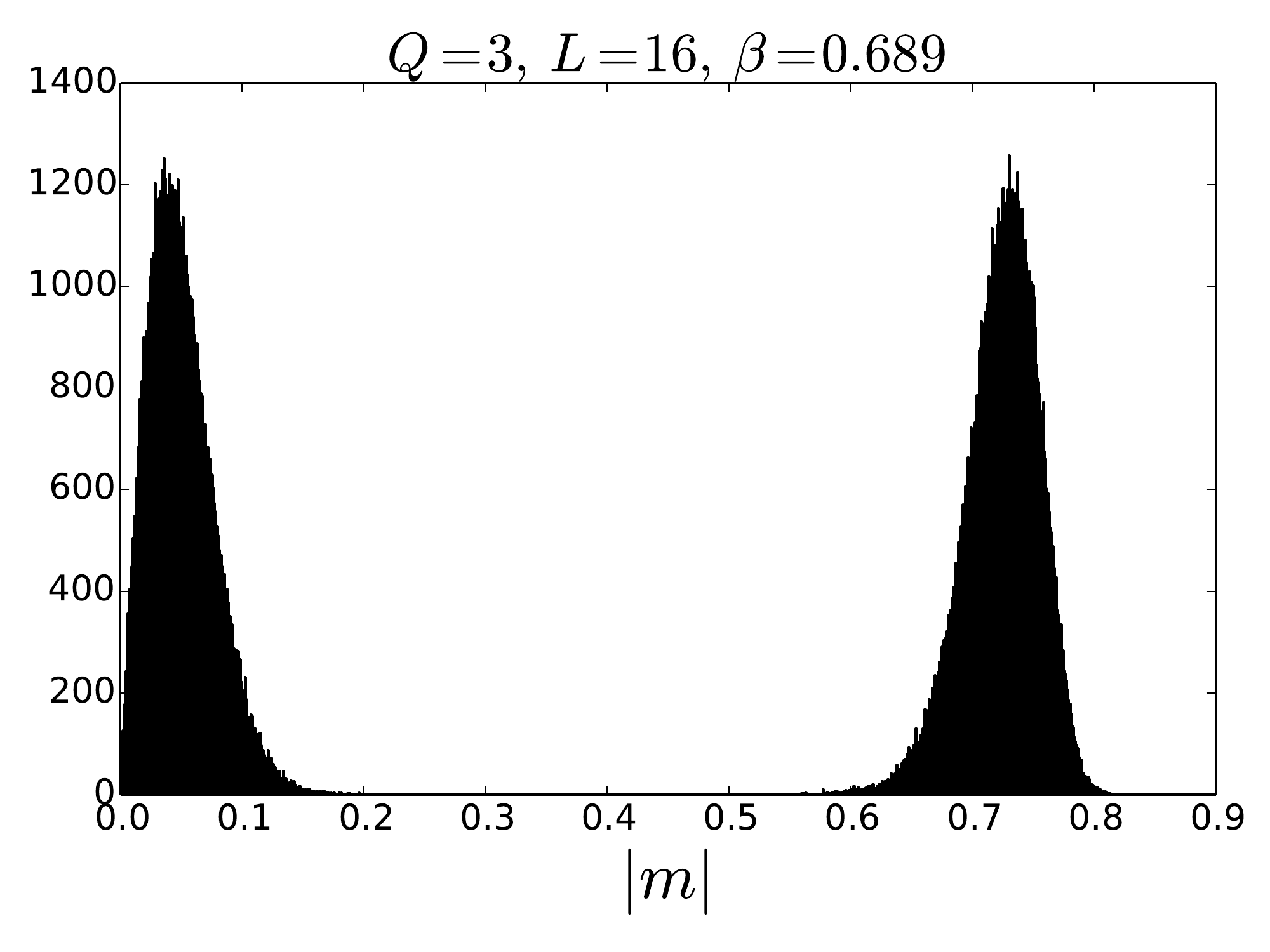}
}
\end{center}\vskip-0.7cm
\caption{Energy density $E$ (top panel) as a function of MC sweep and the histogram of magnetization density $\langle |m| \rangle$ (bottom panel) 
near $T_c$ of the 3-state 
ferromagnetic Potts model on the cubic lattice. The associated box size $L$ 
and $\beta$ for the data shown in this figure are $L = 16$ and $\beta = 0.689$, respectively}
\label{Q53Denergy}
\end{figure}  

\subsection{Results of 3D 5-state ferromagnetic Potts model}

\subsubsection{The Monte Carlo results}
In fig.~\ref{Q53Denergy}, the energy density $E$ as a function 
of MC sweep (top panel),
and the histogram of the magnetization density $\langle |m| \rangle$ 
(bottom panel) at a temperature close to $T_c$ for the
3D 5-state ferromagnetic Potts model on the cubic lattice are shown.
The outcomes are obtained with $L=16$ and $\beta = 0.689$.
The phenomenon of tunneling between two values clearly appear in the top panel of the figure.
In addition, two peaks structure shows up as well in the bottom panel of fig.~\ref{Q53Denergy}.  
These are the features of a first order phase transition. In other words, our Monte Carlo
data confirm the theoretical prediction that the phase transition of 3D 5-state  
ferromagnetic Potts model on the cubic lattice is discontinuous.

\subsubsection{The NN results}

\begin{figure}
\begin{center}
\includegraphics[width=0.35\textwidth]{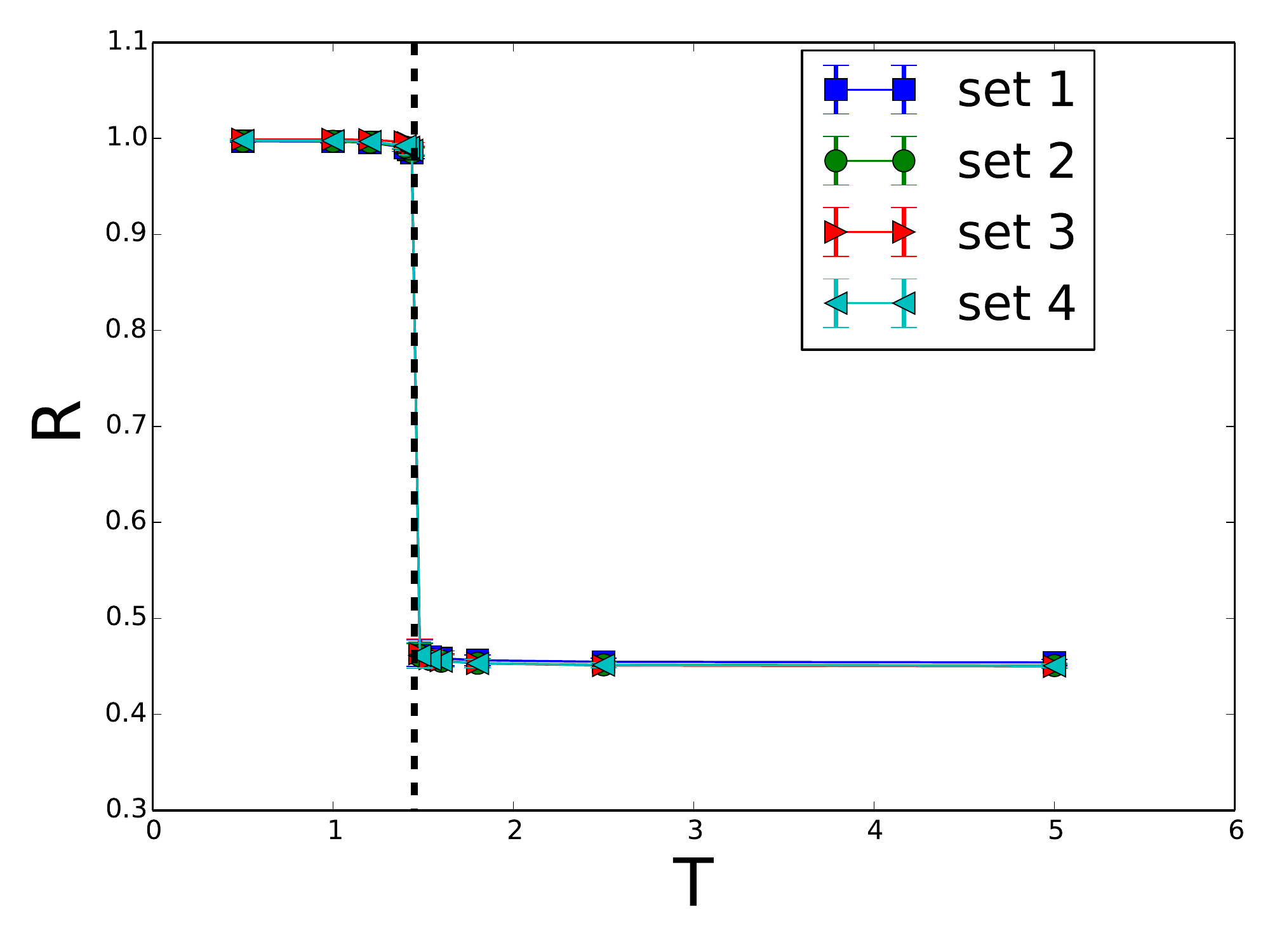}
\end{center}\vskip-0.7cm
\caption{$R$ as functions of $T$ for the 3D 5-state
ferromagnetic Potts model on the cubic lattice with $L=12$. 
The vertical dashed line is the expected $T_c$. Results
obtained using different parameter sets are all shown in the
figure.}
\label{Q53DRT0}
\end{figure}

\begin{figure}
\begin{center}
\includegraphics[width=0.35\textwidth]{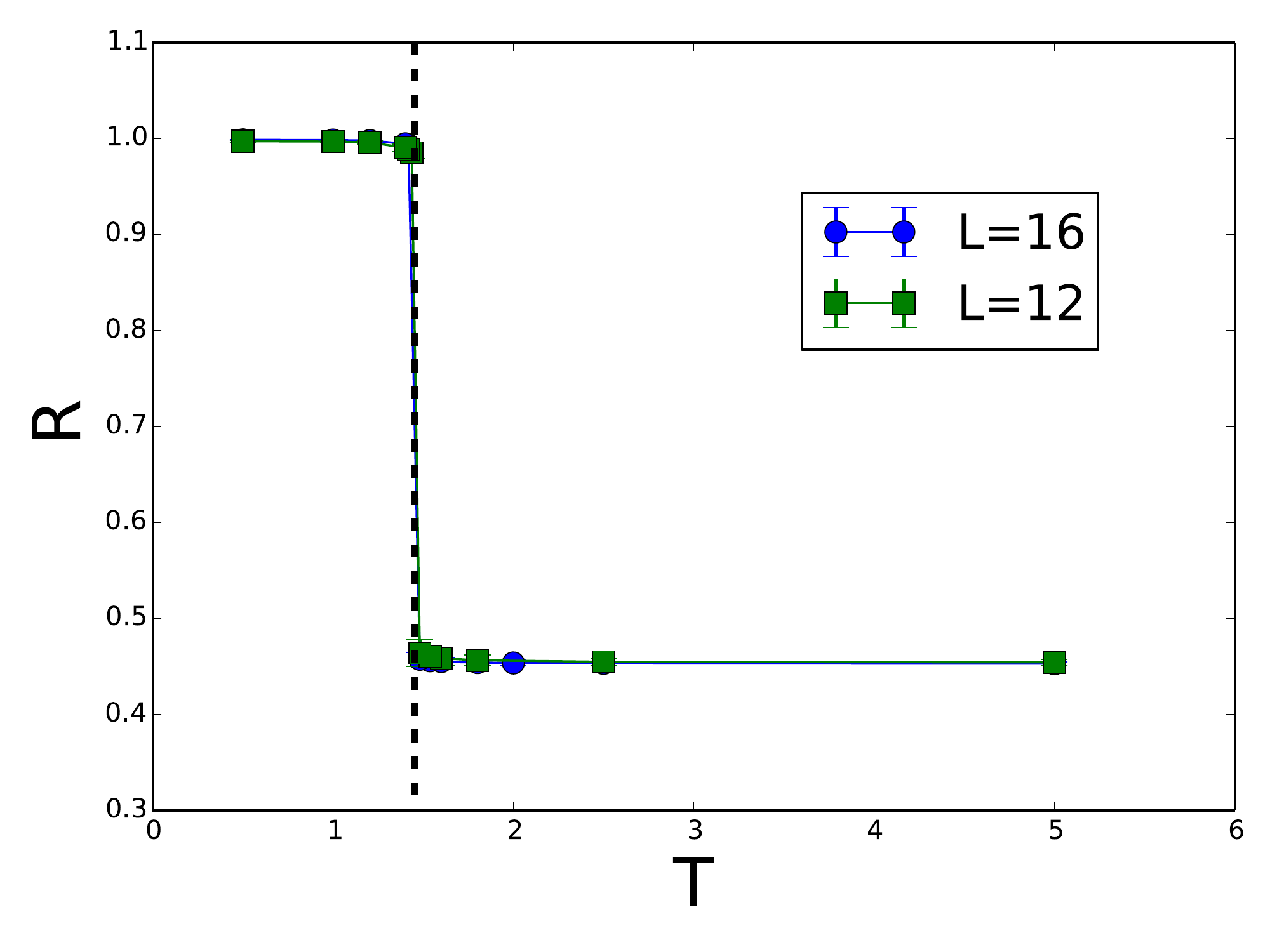}
\end{center}\vskip-0.7cm
\caption{$R$ as functions of $T$ for the 3D 5-state
ferromagnetic Potts model on the cubic lattice with $L=12,16$. 
The vertical dashed line is the expected $T_c$.}
\label{Q53DRT}
\end{figure}

\begin{figure}
\begin{center}
\includegraphics[width=0.35\textwidth]{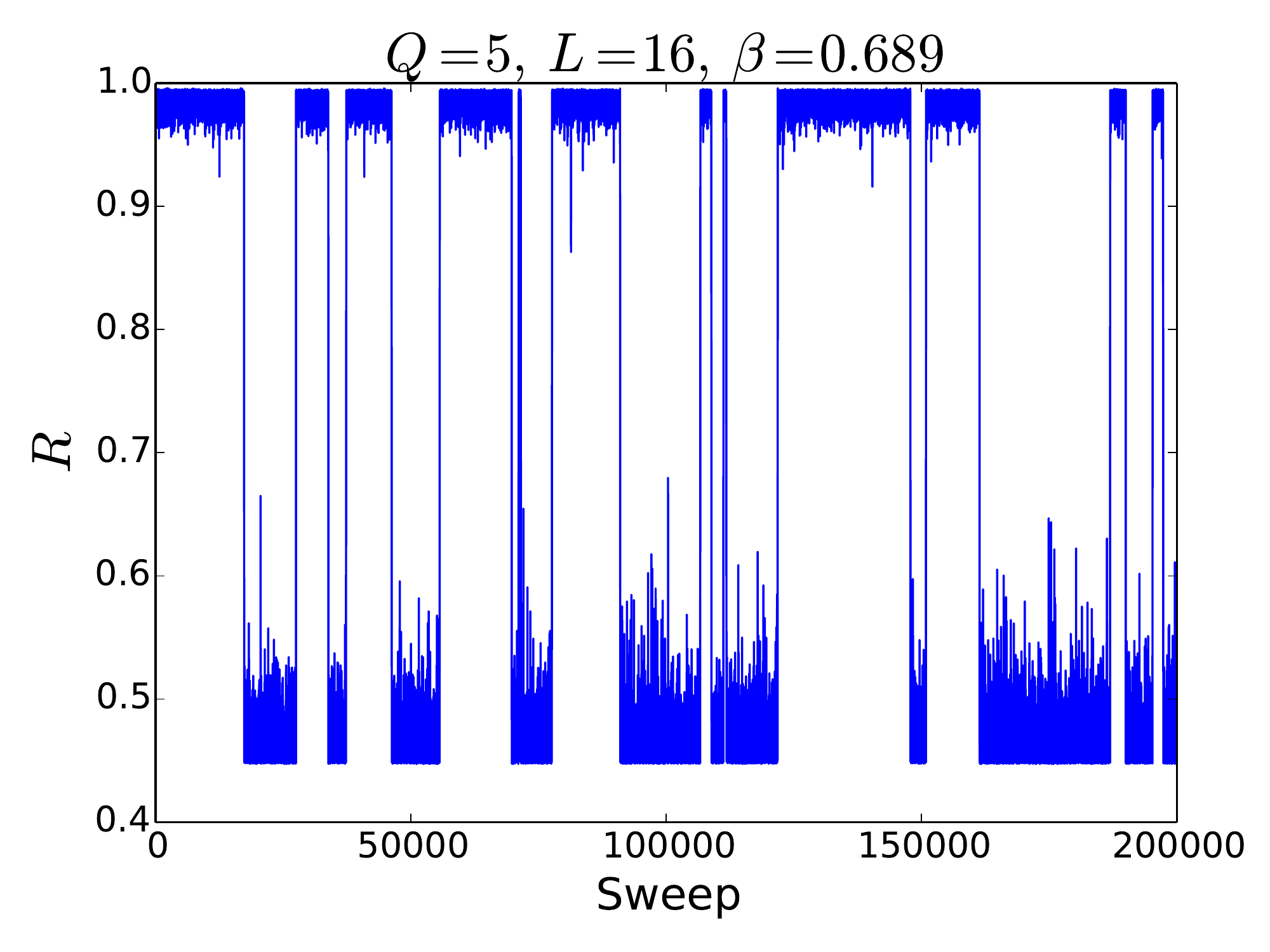}
\end{center}\vskip-0.7cm
\caption{The histogram of $R$ at a temperature near the $T_c$ of 3D 5-state
ferromagnetic Potts model on the cubic lattice. The associated box size $L$
and $\beta$ for the data shown in this figure are $L = 16$ and $\beta = 0.689$, respectively}
\label{Q53DRH}
\end{figure}

The norm $R$ of the output vectors as functions of $T$ 
for the 3D 5-state Potts model on the cubic lattice are 
demonstrated in fig.~\ref{Q53DRT0}. 
The vertical dashed line which appears in
the figure is the expected $T_c$. These results are obtained on
$12$ by $12$ by $12$ lattices. Moreover,
for a fixed $T$ four calculations using different parameters of
random seeds, batchsize, copies of the pre-training set, and epoch are 
conducted and all the obtained
resulting $R$ are shown in fig.~\ref{Q53DRT0}. The outcomes 
in fig.~\ref{Q53DRT0} indicate that
$R$ is very stable with respect to the tunable variables associated with NN. 
In addition, as can be seen from the figure as well as that of 
fig.~\ref{Q53DRT} which includes the outcomes of $L=16$, the 
magnitude of $R$ decreases rapidly in the temperature region close
to the theoretical $T_c$. Based on this result and that of Ref.~\cite{Li18}, it is beyond doubt 
that for $Q$-state ferromagnetic Potts model,
the associated $T_c$ can be precisely estimated
to lie within the temperature window at which the magnitude of $R$ 
drops sharply from 1 to 1/$\sqrt{Q}$   

Fig.~\ref{Q53DRH} shows the histogram of $R$ for the 3D 5-state ferromagnetic
Potts model on the cubic lattice at a temperature $T$ near $T_c$. 
The outcome is obtained with $L=16$ and $\beta = 0.689$.
A clear two peaks structure obviously appears in the figure.
As a result, the studied phase transition is first order.
In other words, the NN constructed here, which consists
of only one input layer, one hidden layer, and one output layer,
is capable of not only locating $T_c$ precisely, but also 
determining the nature of the phase transition of
the investigated model. It is anticipated that the built NN
can carry out similar calculations with success for general
$Q$-state ferromagnetic Potts models in any dimension and on
any lattice geometry.

\subsection{Results of 2D 3-state antiferromagnetic Potts model}

\subsubsection{The Monte Carlo results}

\begin{figure}
\begin{center}
\includegraphics[width=0.35\textwidth]{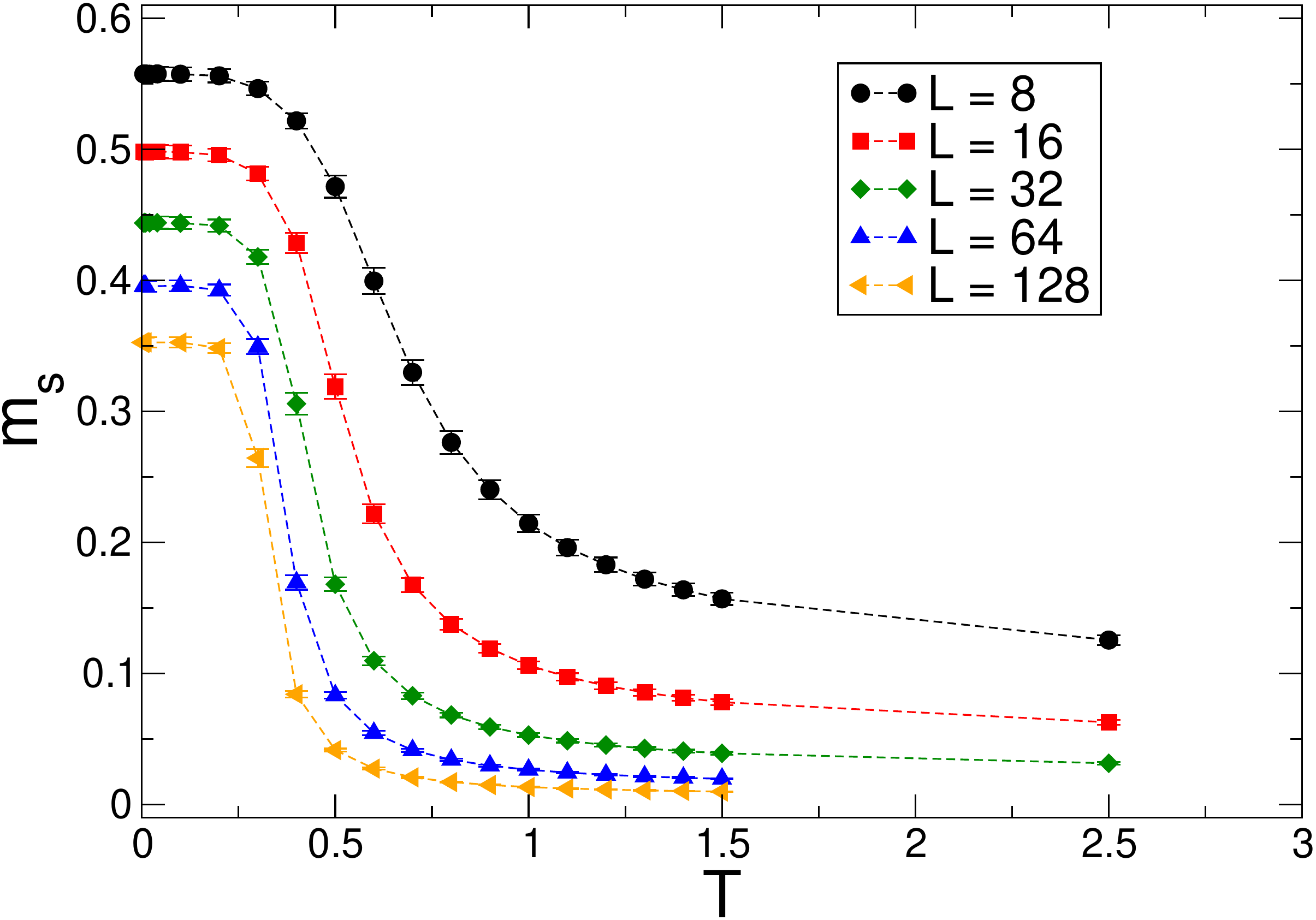}
\end{center}\vskip-0.7cm
\caption{$m_s$ as functions of $T$ for the 2D 3-state antiferromagnetic
Potts model on the square lattice. Results associated with various 
box sizes $L$ are shown in the figure.}
\label{AFMC}
\end{figure}

In fig.~\ref{AFMC} the staggered magnetization density $m_s$ as functions of temperature $T$ 
for the considered
2D 3-state ferromagnetic Potts model on the square lattice are presented. 
In particular, outcomes corresponding to various $L$ are shown in the figure.
The results in the figure demonstrate that for every finite $L$, the 
magnitude of its corresponding magnetization diminishes as $T$ rises and 
eventually at high temperature $m_s$ reaches a saturated value which is 
anticipated to go to zero when $L \rightarrow \infty$. 
Moreover, for the simulated box sizes, the curves shown in the figure do not 
intersect among themselves. Such a scenario
is interpreted as the phase transition takes place at zero temperature, namely 
the system is always in the disordered phase at any $T > 0$.

\subsubsection{The NN results}

\begin{figure}
\begin{center}
\includegraphics[width=0.35\textwidth]{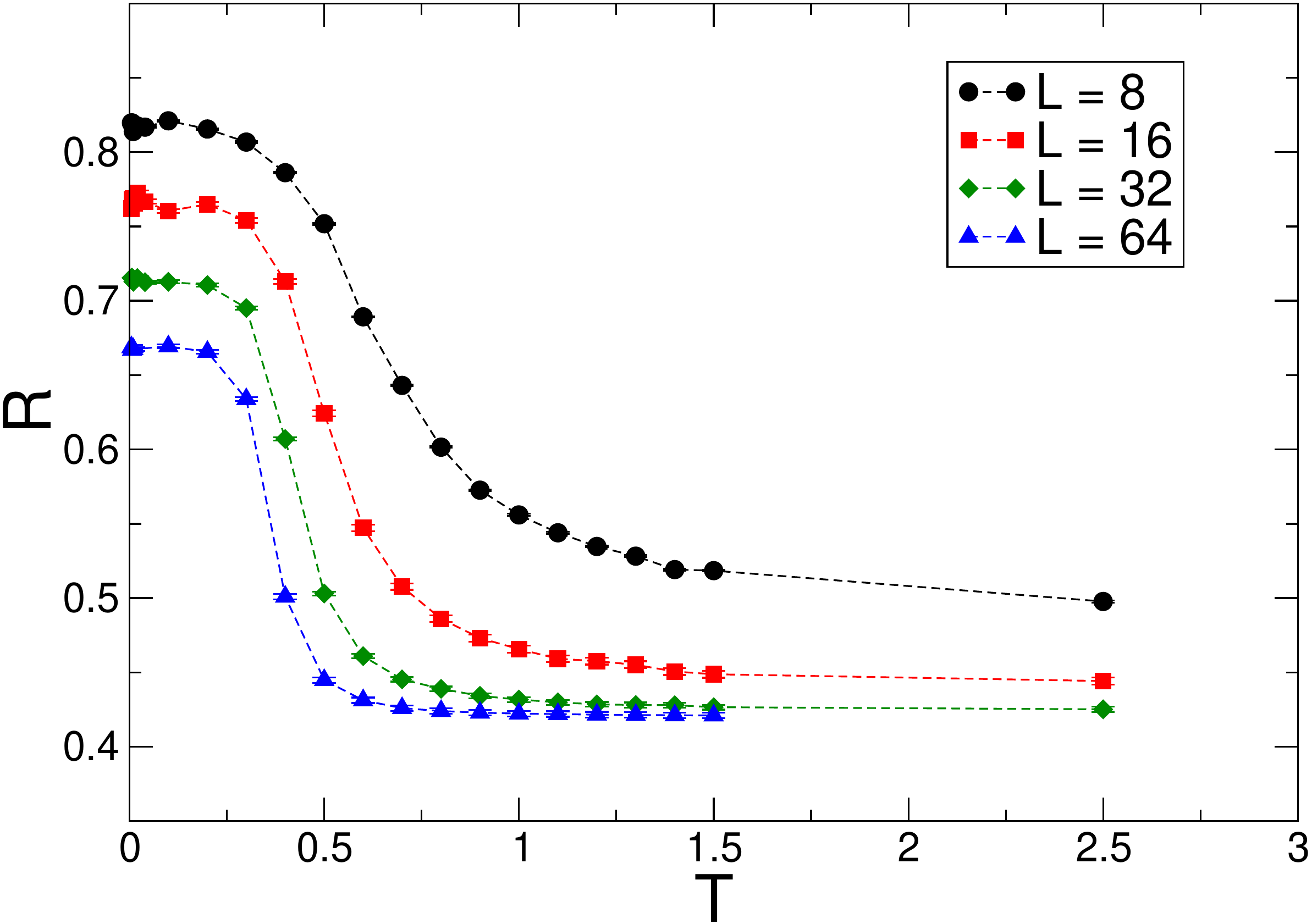}
\end{center}\vskip-0.7cm
\caption{$R$ as function of $T$ for various box sizes $L$. 
These results are obtained using 6 configurations as the pre-training set and the considered batch size is 40.}
\label{R6_1}
\end{figure}

\begin{figure}
\begin{center}
\includegraphics[width=0.35\textwidth]{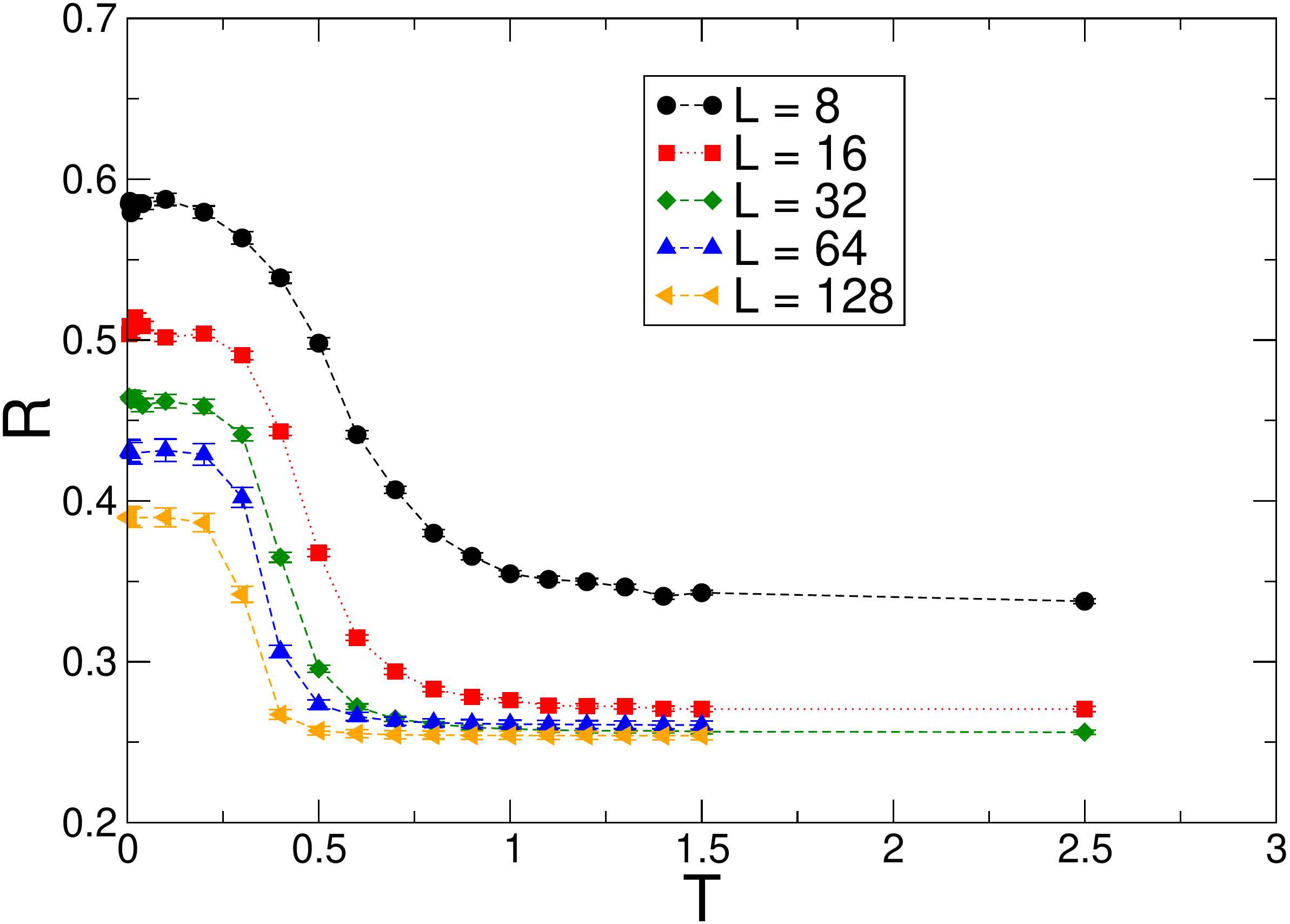}
\end{center}\vskip-0.7cm
\caption{$R$ as functions of $T$ for various box sizes $L$. 
These results are obtained using 18 configurations as the pre-training set and the considered batch size is 320.}
\label{R18_1}
\end{figure}

\begin{figure}
\begin{center}
\includegraphics[width=0.35\textwidth]{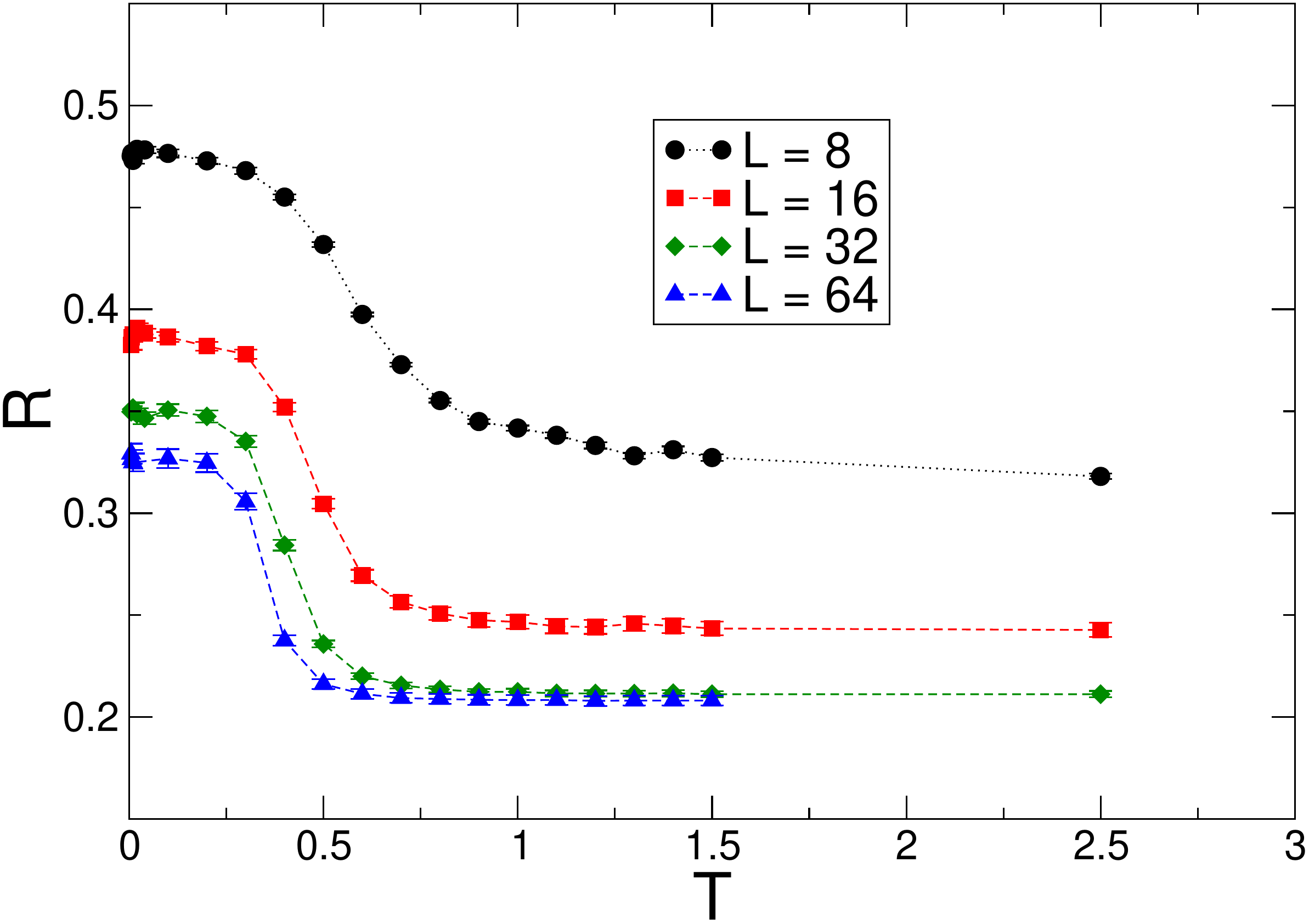}
\end{center}\vskip-0.7cm
\caption{$R$ as functions of $T$ for various box sizes $L$. 
These results are obtained using 36 configurations as the pre-training set and the considered batch size is 320.}
\label{R36_1}
\end{figure}

The NN outcomes of $R$ as functions of $T$ for various training sets (using 6, 18, and 36
constructed configurations as the pre-training sets)
are shown in figs.~\ref{R6_1},~\ref{R18_1},~\ref{R36_1}, and the shown data are 
obtained using
ten results. In particular, each of these results is
calculated with its own set of random seeds which is different from that of the others. 
These figures, which are obtained using different
training sets, all demonstrate similar characteristics as that of
$m_s$ (fig.~\ref{AFMC}). Specifically, the $T$-$R$ curves of various $L$ have the
trend that the magnitude of $R$ decrease monotonically with $T$.
In addition, for every employed training set,
the associate $T$-$R$ curves do not intersect among themselves except
those of larger $L$, which can be interpreted as the size convergence
of the NN outcomes. 

We would like to point out that for the results
of using 18 configurations as the pre-training set
(i.e fig.~\ref{R18_1}), 
in the high temperature region $R$ of $L=64$ are slightly above that of
$L=32$ (not within statistical errors). We attribute this to the 
systematic uncertainty due to the tunable parameters of NN that are not
taken into account here. Nevertheless, considering the similarity between the results
of NN and MC (i.e. figs.~\ref{AFMC},~\ref{R6_1},~\ref{R18_1},~\ref{R36_1}), 
the outcomes of NN provide
convincing numerical evidence that the phase transition of 2D
3-state antiferromagnetic Potts model on the square lattice 
occurs at zero temperature.
 
When compared with that of the studied 3D model,
the NN outcomes of the 2D 3-state antiferromagnetic Potts model
have (much) larger uncertainties. Indeed, for the calculations using various 
random seeds, while the variation among the resulting $R$ associated with the considered 
3D model is negligible, the uncertainty of R related to the studied 2D model
has sizable magnitude. Similarly, other tunable parameters in the used NN such as
the batchsize have certain impact on the outcomes of $R$ of the antiferromagnetic 
system. We further find that in order to obtain results consistent with that 
determined from Monte Carlo simulations, the ratio $p$ between how many objects in
the training set and the considered batchsize has to be a number with moderate magnitude.
Since $p$ is associated with the independent parameters during the optimization
procedure, too many or too few free parameters will lead to 
not satisfactory outcomes from the optimization considering the limitation
of the algorithm employed in this process. 

\begin{figure}
\begin{center}
\vbox{
\includegraphics[width=0.33\textwidth]{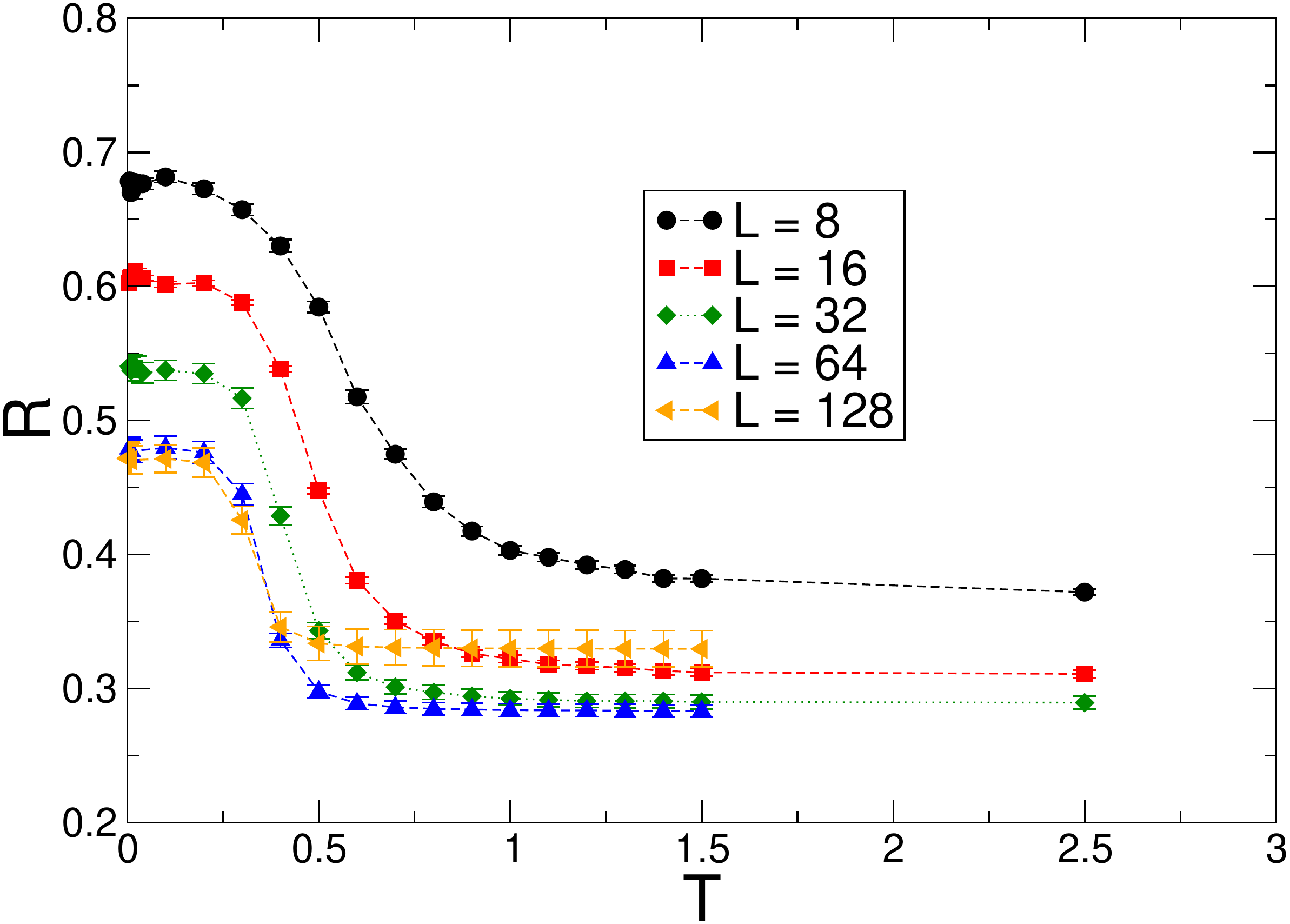}
\includegraphics[width=0.33\textwidth]{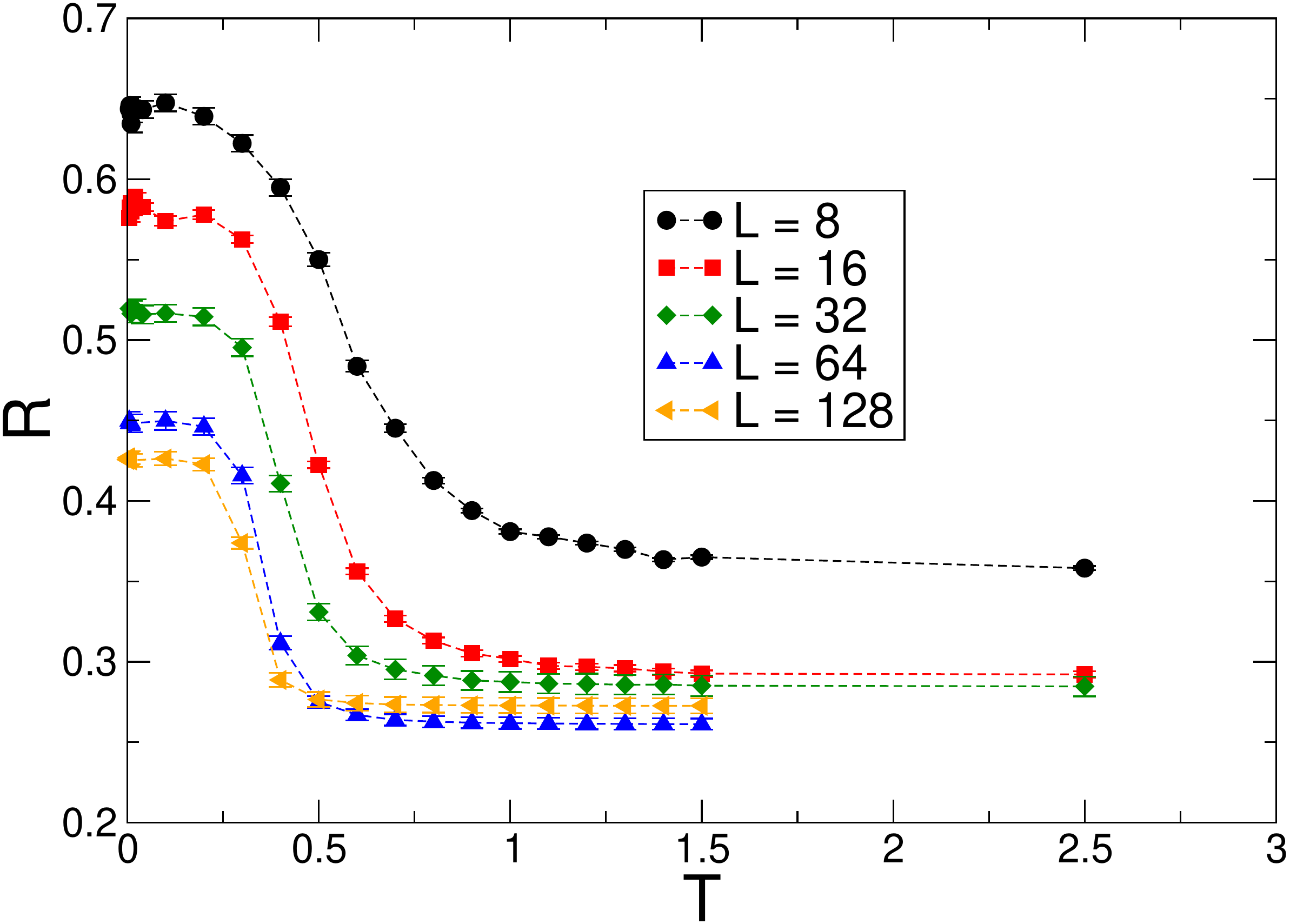}
\includegraphics[width=0.33\textwidth]{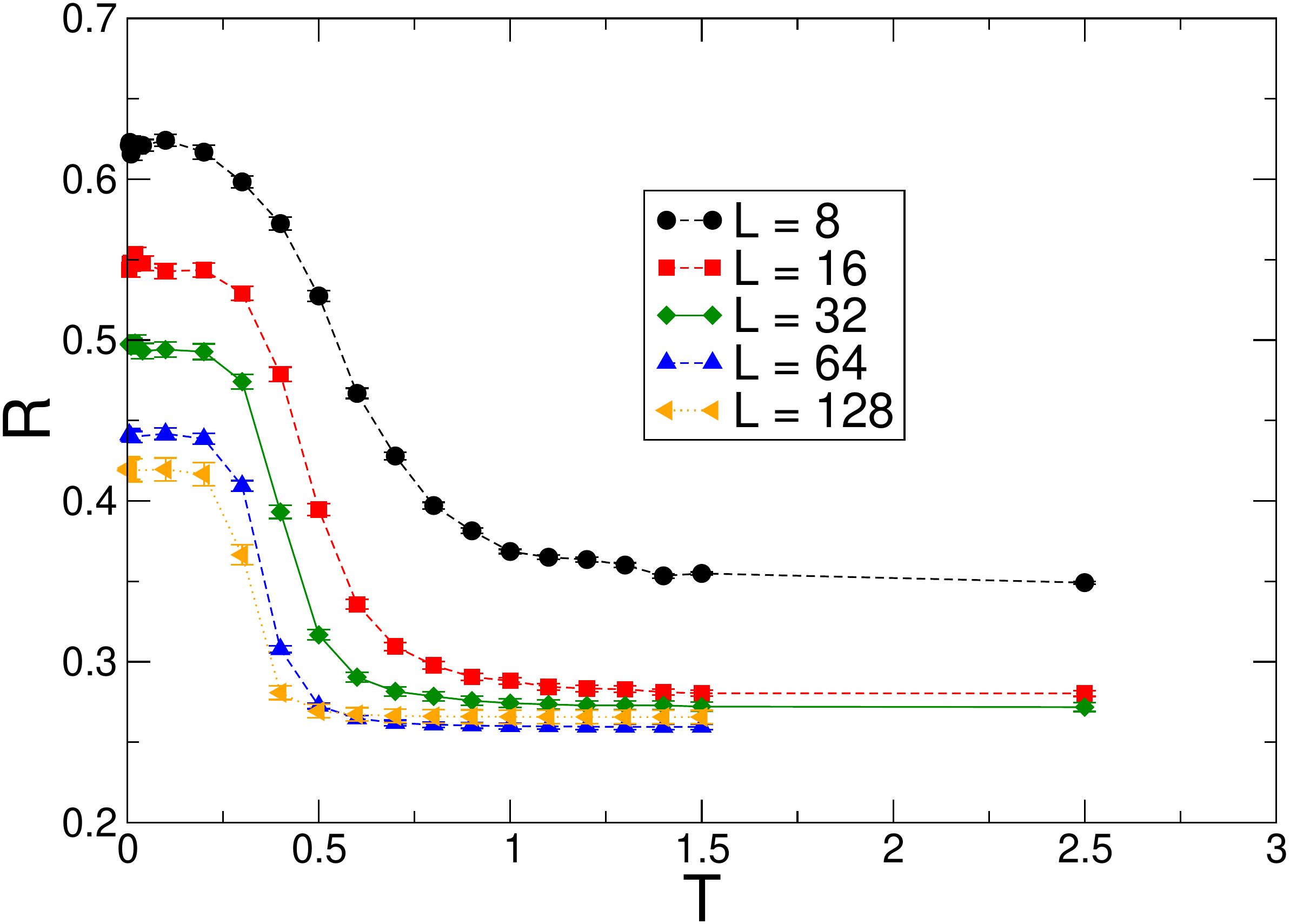}
\includegraphics[width=0.33\textwidth]{C18B320.pdf}
}
\end{center}\vskip-0.7cm
\caption{$R$ as functions of $T$ for various box sizes $L$. 
These results are obtained using 18 configurations as the pre-training set and the employed batch sizes are 
40, 80, 160, and 320 (from top to bottom).}
\label{R18}
\end{figure}

Using 18 configurations as the pre-training set, 
the NN outcomes of $R$ for various $T$ and $L$
obtained using batchsizes 40, 80, 160, and 320
are shown in fig.~\ref{R18} (from top to bottom). As can be seen in that figure,
when batchsize is 40 the corresponding data of $L=128$ lie well above those of $L=32,64$ 
in the high temperature region. This is in contradiction with the Monte Carlo results.   
As the batchsize increases, the trend of $R$ versus $T$ for various 
box sizes $L$ become more and more similar to that of MC. 
Finally, the outcomes shown in the bottom panel of fig.~\ref{R18} which is calculated with batchsize 320
are consistent, at least qualitatively, with that of MC. 
Our investigation of 2D 3-state antiferromagnetic Potts model on the square lattice
indicate that cautions have to be taken, when NN techniques are considered to
study physics systems having highly degenerated ground state configurations.

Finally, we would like to emphasize the fact that the results shown in each of
figs.~\ref{R6_1},~\ref{R18_1},~\ref{R36_1} are obtained using the same variables of 
NN for all the considered
box sizes $L=8,16,32,64 (128)$. It is likely that 
the most suitable parameters associated with NN for various $L$ could be different. 
In other words, for complicated systems, carrying out certain fine tuning to search appropriate
parameters of NN may be required in order to reach the right signals of physics.

\section{Discussions and Conclusions}

In this study we investigate the phase transitions of 3D 5-state ferromagnetic
Potts model and 2D 3-state antiferromagnetic Potts model, using both the
Monte Carlo calculations and techniques of NN. The NN considered 
here has the simplest deep learning structure, namely it consists of one input
layer, one hidden layer, and one output layer. Moreover, unlike the conventional
approach of using data generated by numerical methods for the training,
in our study we employ full or part of the theoretical ground state configurations 
as the (pre-)training sets. 

The conventional training of a NN typically requires the use of actual data points.
In particular, the knowledge of the critical point ($T_c$) is essential to 
study the associated phase transition using the standard approach of 
NN methods. 
Our strategy for the training process has the advantage that 
information of $T_c$ is not necessary to carry out the investigation
and very little computation effort is needed for generating the training sets.  
The magnitude of the output vectors $R$ is shown to be the relevant
quantity to locate the critical points as well as to determine the nature of
the phase transitions.   

Remarkably, the NN results related to the studied 3D models obtained here imply
that even a simplest NN of deep learning can lead to highly accurate 
determination of $T_c$. Furthermore, the quantity $R$ used here is as efficient
as that typically considered in the traditional methods when it comes to decide the nature of the
considered phase transitions. Interestingly, the tunneling phenomena in 
figs.~\ref{Q53Denergy} and \ref{Q53DRH}
indicate that, whenever $E$ reaches the results of large numerical values, 
$R$ obtains the outcomes with small magnitude and vice-versa. In other words, 
$R$ and $E$ are complementary to each other and $R$ indeed reflects the 
correct physics. 

For the 2D 3-state antiferromagnetic Potts model, we have carried out the NN
investigation using 6, 18, and 36 theoretical ground state configurations 
of this model as the (pre-)training sets. While the resulting NN outcomes 
with certain constraints on the tunable parameters are consistent with
the Monte Carlo results, it is subtle to reach the correct physics from the
NN calculations. Indeed, as we have demonstrated here, the variation among NN 
results obtained with different random seeds and batchsize are not 
negligible. In particular, the ratio of the number of training objects and the
batchsize plays a crucial role in obtaining outcomes having the right 
signals of physics. In summary, one has to pay special attention
when models having highly degenerated ground state configurations
are investigated using the NN method. For such cases, 
certain fine tuning to search appropriate
parameters of NN may be required in order to observe the correct physics.  

To conclude, here we reconfirm the validity of the training approach considered 
in Ref.~\cite{Li18}. In particular, we succeed in applying this method to the
study of the phase transition of 2D 3-state antiferromagnetic Potts model on the square 
lattice, which takes place at zero temperature and may be difficult to
detect using the conventional NN training procedure. It remains interesting
to examine whether the method employed in this study is capable of precisely 
calculating other relevant physical quantities at phase transitions, such as 
the critical exponents.

\section*{Acknowledgement}
The first three authors contributed equally to this project. 
Partial support from Ministry of Science and Technology of Taiwan is 
acknowledged.

\end{document}